\documentclass[aps,prb,twocolumn,amsmath,amssymb,superscriptaddress,showpacs]{revtex4-1}

\usepackage[pdftex]{graphicx}
\usepackage{graphicx}
\usepackage{epstopdf}
\usepackage{dcolumn}
\usepackage{bm}
\usepackage{color}
\usepackage[dvipsnames]{xcolor}
\usepackage{epsfig}
\usepackage{latexsym}
\usepackage{amsmath}
\usepackage[normalem]{ulem}
\usepackage{amssymb}
\usepackage{wasysym}
\usepackage{hyperref}
\usepackage{natbib}

\def\kpara{${k}_\parallel$}

\def\kperp{${k}_\perp$}
\def\Gbar{$\overline{\Gamma}$}

\def\BiTe{Bi$_2$Te$_3$}
\def\BiSe{Bi$_2$Se$_3$}

\def\invA{\AA$^{-1}$}

\def\GbarKbar{$\overline{\Gamma}$-$\overline{\rm K}$}

\newcommand{\bise}{Bi$_2$Se$_3$}

\begin{document}

\title{Anomalous behavior of the electronic structure of (Bi$_{1-x}$In$_x$)$_2$Se$_3$ across the quantum-phase transition from topological to trivial insulator}

\author{J. S\'anchez-Barriga}
 \affiliation{Helmholtz-Zentrum Berlin, Albert-Einstein-Str. 15, D-12489 Berlin, Germany\\}

\author{I. Aguilera}
\affiliation{Peter Gr{\"{u}}nberg Institute and Institute for Advanced Simulation, Forschungszentrum J{\"{u}}lich and JARA, D-52425 J{\"{u}}lich, Germany}

\author{L. V. Yashina}
 \affiliation{Department of Chemistry, Moscow State University, Leninskie Gory 1/3, 119991 Moscow, Russia\\}

\author{D. Y. Tsukanova}
 \affiliation{Department of Chemistry, Moscow State University, Leninskie Gory 1/3, 119991 Moscow, Russia\\}

\author{F. Freyse}
 \affiliation{Helmholtz-Zentrum Berlin, Albert-Einstein-Str. 15, D-12489 Berlin, Germany\\}

\author{A. N. Chaika}
\affiliation{Institute of Solid State Physics RAS, Academician Ossipyan str. 2, Chernogolovka, 142432 Moscow District, Russia\\}

\author{A. M. Abakumov}
 \affiliation{Skolkovo Institute of Science and Technology, 3 Nobel Street, 143026 Moscow, Russia\\}

\author{A. Varykhalov}
 \affiliation{Helmholtz-Zentrum Berlin, Albert-Einstein-Str. 15, D-12489 Berlin, Germany\\}

\author{E. D. L. Rienks}
 \affiliation{Helmholtz-Zentrum Berlin, Albert-Einstein-Str. 15, D-12489 Berlin, Germany\\}

\author{G. Bihlmayer}
 \affiliation{Peter Gr{\"{u}}nberg Institute and Institute for Advanced Simulation, Forschungszentrum J{\"{u}}lich and JARA, D-52425 J{\"{u}}lich, Germany}

\author{S. Bl\"{u}gel}
 \affiliation{Peter Gr{\"{u}}nberg Institute and Institute for Advanced Simulation, Forschungszentrum J{\"{u}}lich and JARA, D-52425 J{\"{u}}lich, Germany}

\author{O. Rader}
 \affiliation{Helmholtz-Zentrum Berlin, Albert-Einstein-Str. 15, D-12489 Berlin, Germany\\}


\begin{abstract}
Using spin- and angle-resolved spectroscopy and relativistic many-body calculations, we investigate the evolution of the electronic structure of (Bi$_{1-x}$In$_x$)$_2$Se$_3$ bulk single crystals around the critical point of the trivial to topological insulator quantum-phase transition. By increasing $x$, we observe how a surface gap opens at the Dirac point of the initially gapless topological surface state of Bi$_2$Se$_3$, leading to
the existence of massive fermions. The surface gap monotonically increases for a wide range of $x$ values across the topological and trivial sides of the quantum-phase transition. By means of photon-energy dependent measurements, we demonstrate that the gapped surface state survives the inversion of the bulk bands which occurs at a critical point near $x=0.055$. The surface state exhibits a non-zero in-plane spin polarization which decays exponentially with increasing $x$, and that persists on both the topological and trivial insulator phases. Its out-of-plane spin polarization remains zero demonstrating the absence of a hedgehog spin texture expected from broken time-reversal symmetry. Our calculations reveal qualitative agreement with the experimental results all across the quantum-phase transition upon the systematic variation of the spin-orbit coupling strength. A non-time reversal symmetry breaking mechanism of bulk-mediated scattering processes that increase with decreasing spin-orbit coupling strength is proposed as explanation.
\end{abstract}

\maketitle

\section{Introduction}

Three-dimensional (3D) topological insulators (TIs) are characterized by an insulating bulk band gap and Dirac cone topological surface states (TSSs). Band inversion causes the bulk band gap and depends strongly on spin-orbit coupling (SOC). The surface states are the result of the topological properties of the bulk via a bulk-boundary correspondence. They are protected by time-reversal symmetry and robust against disorder, for example by impurity scattering \cite{FuKane07,Moore10,HasanKane10,Qi11}. The high SOC causes a linear quasirelativistic band dispersion, a Dirac crossing point at time-reversal invariant momenta, and a spin texture in momentum space which locks spin and linear momentum \cite{Hsieh2009,Pan11,Jozwiak11, Sanchez-Barriga-PRX-2014}. 

The bulk-boundary correspondence creates topologically protected states at the surface and also at interfaces with topologically-trivial matter. This fact requires consideration when investigating the topological phase transition between a topologically protected and a trivial bulk phase. For example, the influence of magnetic impurities has previously been investigated, both at the surface and in the bulk of prototypical TIs such as \BiSe\ and \BiTe. At the surface, magnetic impurities that are not magnetically aligned perpendicular to the surface do not open a band gap at the Dirac point of the TSS up to thicknesses of at least a monoatomic layer \cite{Valla12,Scholz12,Scholz13,Honolka12,Sessi14}.  

In contrast, magnetic impurities in the bulk can open large band gaps of the order of 100 meV in the case of Mn in \BiSe\cite{Sanchez16}. These band gaps were found not to be of magnetic origin \cite{Sanchez16} because they persist for temperatures high above the Curie temperature. Indeed, the magnetic anisotropy of Mn in \BiSe\ was found to be in the surface plane, and the size of the gap at the Dirac point exceeds band structure predictions \cite{Schmidt11,Henk12,Abdalla13} by an order of magnitude. Moreover, angle-resolved photoemission (ARPES) experiments have shown that Mn doping leaves the inverted bulk band structure of \BiSe\ unaffected \cite{Sanchez16}.

In other cases, however, nonmagnetic impurities give rise to a gapped Dirac cone on the surface while the same atomic species induce a topological phase transition in the bulk \cite{wu2013,xu2015}. This has been reported for substitutional In in \BiSe\ where a transition to a trivial insulator was found by transport and ARPES for In$_2$Se$_3$ concentrations between 3 to 7 mol.\% in films of 60 quintuple layers (QLs) thickness \cite{brahlek2012}. Subsequently, the same system was investigated by THz phototransport as a function of thickness and composition \cite{wu2013}. It was found that the critical concentration slightly decreased from 6 to 4 mol.\%\ with decreasing film thickness \cite{wu2013}.

The case of In in \BiSe\ has previously been studied for thin films where the surface-interface coupling was invoked to explain the surface band gap \cite{brahlek2012,wu2013}. It is not established whether this interpretation applies to bulk samples as well. In fact, the first work on (Bi$_{1-x}$In$_x$)$_2$Se$_3$ as bulk samples appeared recently\cite{lou2015}, and reported that the bulk band gap stays negative and constant from about $x$=0.025 to 0.1 and then reverts suddenly, a behavior different from the one reported previously. In a similar context, a surface gap opening has also been reported for Pb$_{1-x}$Sn$_x$Se across the phase transition from topological crystalline insulator to trivial insulator as a function of temperature \cite{Wojek13}. It was also shown that Bi-doping of Pb$_{1-x}$Sn$_x$Se epilayers opens a surface gap at the Dirac cone located at the \Gbar\ point of the surface Brillouin zone (SBZ), inducing a quantum-phase transition from a topological crystalline insulator to a Z$_2$ TI \cite{Mandal2017}.

There is another system that should be mentioned where the behavior of the TSS has been investigated on different sides of the phase transition. In the transition from topological TlBiSe$_2$ to trivial TlBiS$_2$, a gapped surface state was observed on the topological side of the phase transition \cite{Sato11} with reduced spin polarization \cite{Souma12}, as well as on the trivial side \cite{xu2015}. However, the underlying mechanism responsible for the opening of the surface gap has remained elusive so far. All these observations call for a deeper understanding of the unique behavior of surface states across topological phase transitions.

Therefore, in the present work, using (Bi$_{1-x}$In$_x$)$_2$Se$_3$ single crystals as an exemplary system, we perform high-resolution ARPES and spin-resolved ARPES experiments to systematically investigate the evolution of the electronic structure across the trivial to 3D TI quantum-phase transition. By comparing the experimental results to relativistic many-body calculations, we find good qualitative agreement all across the quantum-phase transition upon systematic variation of the In content and the SOC strength, respectively. This concerns both the changes observed in the spin polarization as well as the opening of a gap in the surface states on both sides of the phase transition. In particular, we conclude that a non-time reversal symmetry breaking mechanism of bulk-mediated scattering processes that increase with decreasing SOC strength underlies the unconventional behavior of the surface states. Our results are of general importance in the context of quantum-phase transitions from topological to trivial insulators.

\section{Methods}

Experiments were performed on (Bi$_{1-x}$In$_x$)$_2$Se$_3$ single crystals cleaved {\it in situ} and grown by the Bridgman method \cite{Shtanov2009}. 
The crystals were grown using different compositions from the melt. The growth temperatures were in the range of 650--700$^{\circ}$C, and the temperature gradient during growth was 5$^{\circ}$C/cm. Their composition was characterized by x-ray fluorescence analyisis, energy-dispersive x-ray analysis (EDX) and x-ray diffraction. The high crystal quality of the obtained (111) surfaces was verified by low-energy electron diffraction and by the presence of sharp features in the ARPES dispersions.

Photoemission experiments were carried out using linearly-polarized undulator radiation at the UE112-PGM2 and U125/2-SGM beamlines of the synchrotron source BESSY-II in Berlin. Photoelectrons were detected with a Scienta R8000 electron analyzer and the base pressure of the experimental setup was better than $1\times10^{-10}$ mbar. For spin analysis, a Rice University Mott-type spin
polarimeter \cite{Burnett1994} capable of detecting both in-plane and out-of-plane components of the spin polarization was operated at 26 kV. Overall resolutions of ARPES measurements were 20 meV (energy) and 0.3$^{\circ}$ (angular). Resolutions of spin-resolved ARPES were 80 meV and 1.5$^{\circ}$. 

Scanning tunneling microscopy (STM) experiments were conducted with polycrystalline W tips using a room temperature microscope GPI-300 at pressures below $8\times10^{-11}$ mbar. For high-angle annular dark field scanning transmission microscopy (HAADF-STEM) imaging and EDX mapping, cross-sectional samples were prepared on Cu and Be supports by focused ion-beam milling. A carbon layer was deposited onto the material to protect it from damage during preparation. The HAADF-STEM and EDX data were acquired on aberration-corrected FEI Titan transmission electron microscopes operating at 200 kV, one of which is equipped with a Super-X detector.

The relativistic many-body $GW$ calculations~\footnote{For the $GW$ calculations, we used an angular momentum cutoff of $l=5$ and a linear momentum cutoff of 2.9~bohr$^{-1}$. We included 500 bands and semicore \emph{d} states of Bi and Te. An 8$\times$8$\times$8 \textbf{k}-point mesh was employed.} of the bulk systems were performed with the all-electron full-potential linearized augmented-plane-wave (FLAPW) code {\sc spex}{}~\cite{friedrich2010-spex}. To systematically vary the SOC strength, we took the SOC already into account in the reference system~\cite{sakuma2011,aguilera2013-2} instead of as \textit{a posteriori} correction. The local-density approximation (LDA), as implemented in the code {\sc fleur}{}~\cite{fleur}, was used as a starting point~\footnote{For the LDA calculations, we employed an angular momentum cutoff for the 
atomic spheres of $l=10$ and a plane-wave cutoff in the interstitial
region of 4.5~bohr$^{-1}$. The SOC was incorporated self-consistently employing the second-variation technique.}. As in Ref.~\onlinecite{aguilera2013}, we solved the $GW$ quasiparticle equation in the basis of the LDA single-particle states explicitly. This takes the off-diagonal elements of the self-energy into account, allowing for changes in the quasiparticle wave functions. This is critical to obtain reliable values of the surface-state localization and spin polarization at a $GW$ level. We used the experimental lattice structure of Ref.~\onlinecite{perez1999}. To simulate the surface states, a tight-binding supercell Hamiltonian was constructed from the bulk $GW$-based matrix elements using maximally localized Wannier functions obtained by the {\sc wannier90}{} library~\cite{wannier90}.

\section{Results and discussion}

\begin{figure}[tb!]
\centering
\includegraphics[width=0.42\textwidth]{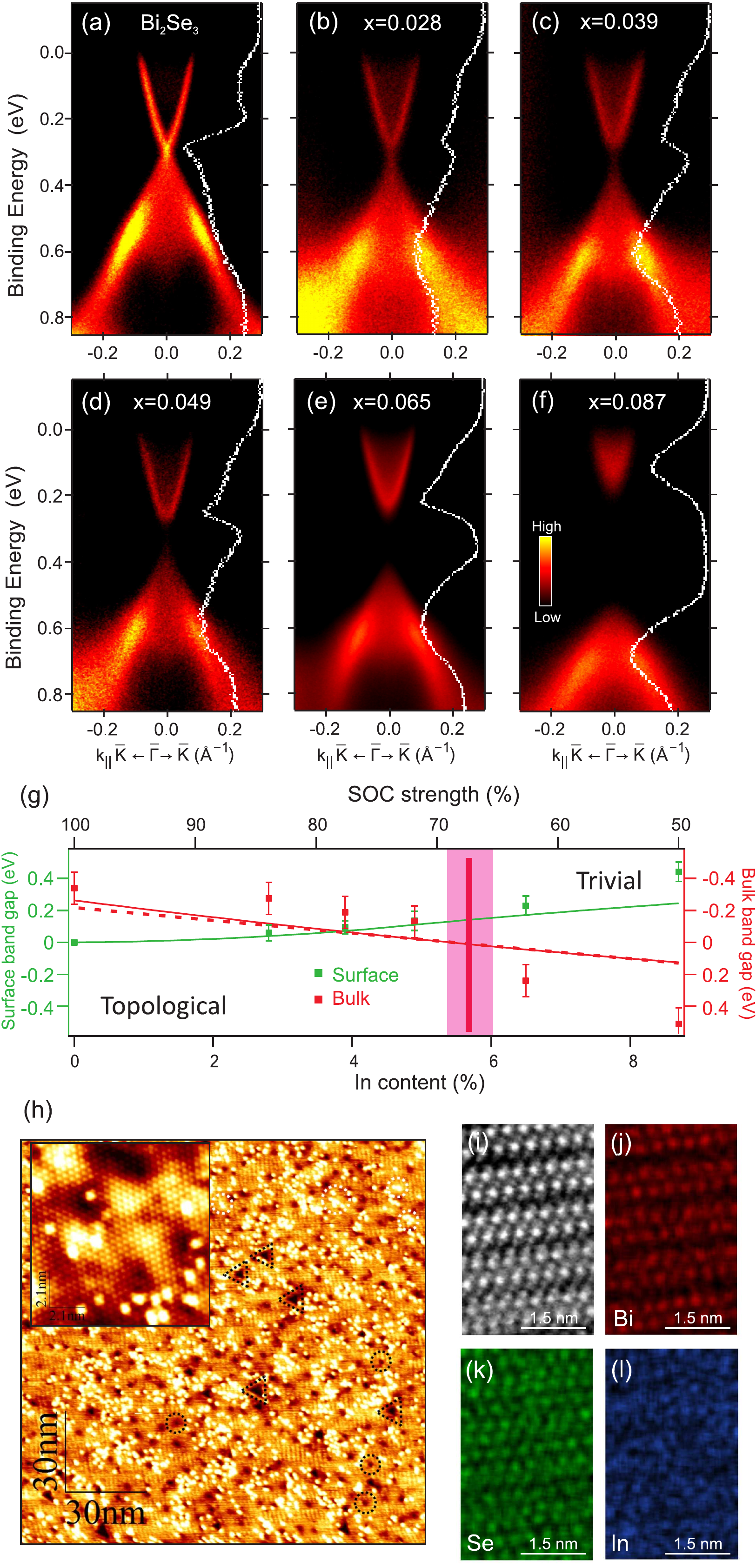}
\caption{\label{fig1expt}(Color online) (a)-(f) High-resolution ARPES dispersions (h$\nu$=50 eV) for $x$ ranging from (a) 0 to (f) 0.087. The vertical white spectra on each panel represent EDCs obtained in normal emission (\kpara=0). (g) Size of the band gaps relative to the transition point. Red (green) solid symbols are the experimental bulk (surface) band gap, and solid lines the corresponding calculations referenced to the SOC~strength. The red-dashed line is the bulk band gap as extracted from photoluminescence experiments. The region of the transition point is emphasized in red. (h) STM image ($x$=0.049) taken at U=-0.8 V and I=60 pA revealing three different types of In-related defects, protrusions and depressions (triangular or circular). Inset: Atomically-resolved STM image taken at U=-20 mV and I=57 pA. (i) High-resolution HAADF-TEM image (x=0.049), and individual (j) Bi, (k) Se and (l) In atomic EDX elemental maps.}
\end{figure}

Figures~\ref{fig1expt}(a)-\ref{fig1expt}(f) show the evolution of electronic band structure of (Bi$_{1-x}$In$_x$)$_2$Se$_3$ around the \Gbar\ point of the SBZ with increasing In content. The high-resolution ARPES dispersions were taken along the \GbarKbar\ direction of the SBZ, and at a photon energy of h$\nu$=50 eV. At this photon energy, contributions from the bulk-conduction band (BCB), which is partially occupied, do not appear due to the dependence of the photoemission transitions on the component of the electron wave vector perpendicular to the surface ${k}_\perp$ \cite{Hufner2007}. While the TSS of \BiSe\ in Fig.~\ref{fig1expt}(a) exhibits a well-resolved and intense Dirac point, in Figs.~\ref{fig1expt}(b)-\ref{fig1expt}(f) we clearly observe how a band gap opens in the surface state with increasing In content. The opening of the surface gap is also evident from the corresponding energy-distribution curves (EDCs) at zero momentum shown on the right-hand side of each panel. While the pronounced Dirac point of \BiSe\ is seen as an intense peak in the EDC of Fig.~\ref{fig1expt}(a), an intensity dip develops at the binding energy of the original Dirac point in the EDCs of Figs.~\ref{fig1expt}(b)-\ref{fig1expt}(f).  
The surface gap can be even observed for the smallest In content and rapidly exceeds 100 meV for $x=0.049$, i.e, it opens in a wide range of $x$ values before the bulk band gap closes [Fig.~\ref{fig1expt}(g)]. This behavior is consistent with the existence of a gapped surface state on the topological side of the phase transition, as also supported by the evolution of the bulk band gap according to photoluminescence data \cite{Galeeva2016} on the same samples [red-dashed line in Fig.~\ref{fig1expt}(g)] and photon-energy dependent ARPES measurements which will be discussed in more detail below [red symbols in Fig.~\ref{fig1expt}(g)]. The opening of the surface gap is also consistent with the fact that the TSS dispersion becomes slightly parabolic near the border of the gap, causing already for $x=0.049$ a small relative increase in the effective mass by $\Delta m^*\approx 0.011 m_e$ with respect to Bi$_2$Se$_3$, where $m_e$ is the free-electron mass.

The structural data of (Bi$_{1-x}$In$_x$)$_2$Se$_3$ are summarized in  Figs.~\ref{fig1expt}(h)-\ref{fig1expt}(l). It is known that for the Bi$_2$Se$_3$-In$_2$Se$_3$ system the in-plane lattice constant $a$ obeys Vegard's law \cite{Lostak1990}, evidencing substitutional behavior in the cation sublattice. From this fact one can expect that major part or even all In atoms occupy cation positions. Combining cross-sectional HAADF-TEM with high resolution EDX elemental maps and STM of the sample surface obtained by cleavage along the van der Waals (vdW) gap is a powerful tool to understand the crystal structure at the atomic level. The transmission microscopy images in Figs.~\ref{fig1expt}(i)-\ref{fig1expt}(l) indicate that mixed crystals of (Bi,In)$_2$Se$_3$ possess a regular structure composed of QLs. From the EDX maps it is clear that the structure of each QL is Se-Bi-Se-Bi-Se, where In is preferentially found within QLs, and in much less amount in the vdW gaps. Nevertheless, the STM image of the cleaved surface in Fig.~\ref{fig1expt}(h) demonstrates a number of structural features or electronic irregularities like white spots, dark triangles and small circles (some of them indicated by dotted lines). The white spots or round shaped bright features, which are visible only at negative bias, are due to In adatoms. These adatoms are seen because of their original location in the vdW gap. Note that their amount is overestimated since they are very mobile and collected by the tip, especially at negative bias, as it was previously shown for In adatoms on Si surfaces \cite{Saranin1999,Sakamoto2003}. Therefore, this observation, in general, evidences the presence of some In atoms in the vdW gap. 
\begin{figure*}[tb!]
\centering
\includegraphics [width=0.8\textwidth]{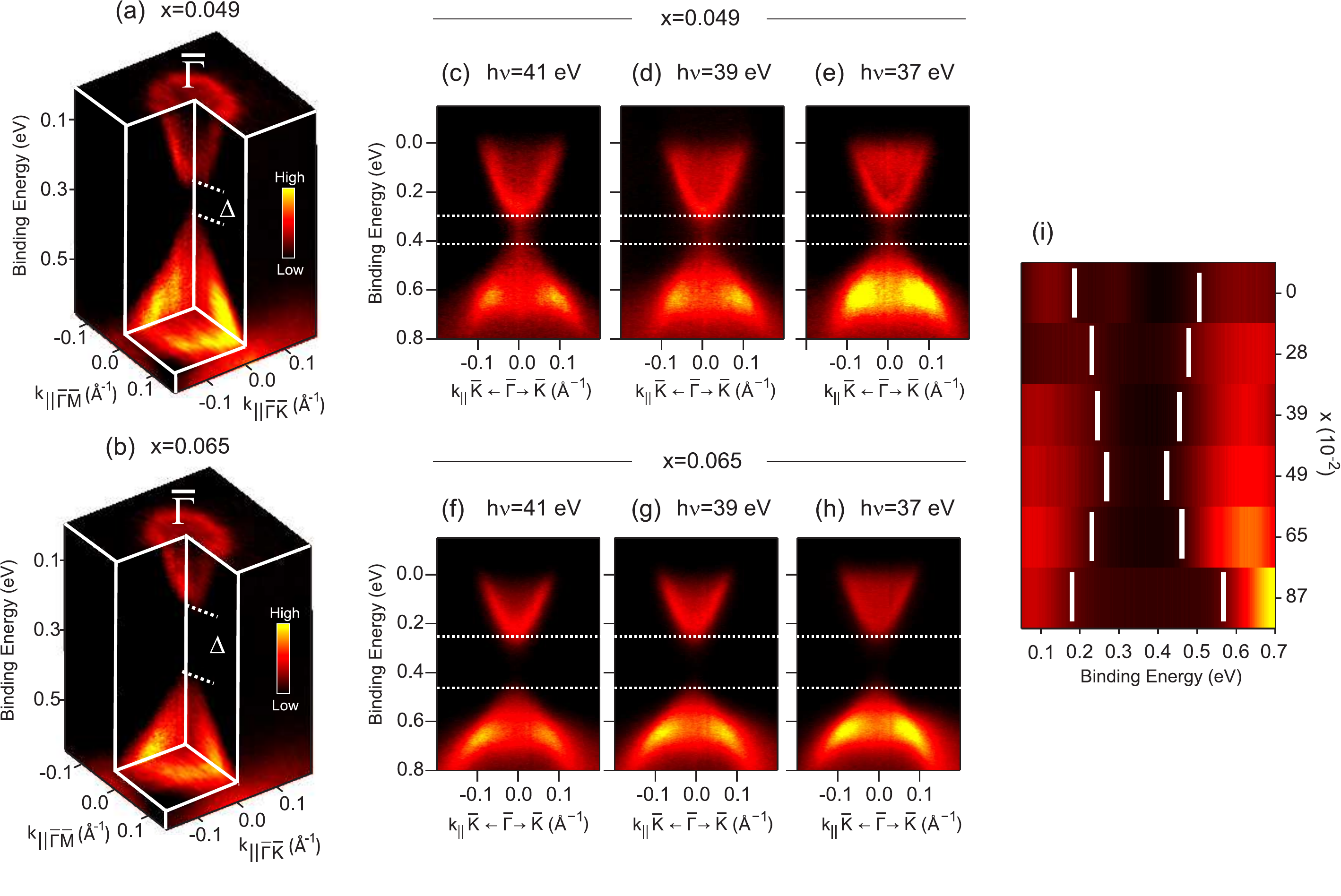}
\caption{\label{hvdep}(Color online) (a),(b) Full ARPES maps obtained with 50 eV photons around the \Gbar\ point of the SBZ for (a) $x$=0.049 and (b) $x$=0.065. In each case, the surface gap $\Delta$ is also indicated. (c)-(h) Photon-energy dependence of the ARPES dispersions, indicating that the gapped surface states are two-dimensional. This is clearly seen at photon energies above 37 eV where the intensity of the surface states is enhanced with respect to the one of bulk states. The surface gap is highlighted by horizontal dashed lines [(c)-(e) $x$=0.049 and (f)-(h) $x$=0.065]. (i) Dependence of the bulk band gap with In content. The data were obtained after sampling the dispersion of the bulk bands as a function of \kperp\ for photon energies below 36 eV, where the intensity of the bulk bands is considerably enhanced. The resulting dispersions were then integrated over \kperp\ around \kpara=0. The evolution of the (direct) bulk band gap is indicated by vertical solid lines, which are a guides to the eye.}
\end{figure*} 

The dark spots, on the other hand, are due to substitutional In atoms at the Bi positions within the subsurface layer and, correspondingly, triangles with lateral dimensions of several nanometers are In$_2$Se$_3$ clusters caused by In atoms in the cation sublattice. These features are well resolved at large bias voltages, as Fig.~\ref{fig1expt}(h) illustrates. However, atomically resolved-patterns measured at small bias voltages and tunneling gaps [inset of Fig.~\ref{fig1expt}(h)] reveal a regular hexagonal lattice proving very low number of defects in the top Se layer. Overall, In atoms prefer substitutional positions in the cation sublattice with notable tendency to form clusters there. This finding is in line with the equilibrium phase diagram of the Bi$_2$Se$_3$-In$_2$Se$_3$ system \cite{Bouanani1996}, which shows thermodynamic tendency to demixing. 

The surface band gap, which is perfectly reproducible for the different samples and from cleavage-to-cleavage, can also be seen in full photoemission mapping on the topological [Fig.~\ref{hvdep}(a)] and trivial [Fig.~\ref{hvdep}(b)] sides of the phase transition. Figures~\ref{hvdep}(c)-\ref{hvdep}(h) demonstrate that the gapped surface states do not disperse with photon energy, i.e., momentum perpendicular to the surface plane \kperp, pinpointing their two-dimensional nature. This behavior, which contrasts the case of bulk states which are known to be highly dispersive with \kperp, is most clearly seen at photon energies above 37 eV where the intensity of the surface states is enhanced with respect to the one of bulk states. From their lack of \kperp\ dispersion we thus conclude that the gapped surface states survive the inversion of the bulk bands. This result is completely different than the common expectation in which surface states are abruptly destroyed when the system moves to the trivial phase due to the change of topology \cite{wu2013,xu2015,brahlek2012}. 

The persistence of gapped surface states in both the topological and trivial phases is further supported by the evolution of the (direct) bulk band gap as a function of In content [see Fig.~\ref{hvdep}(i)]. The data were obtained after sampling the dispersion of the bulk bands as a function of \kperp\ in very fine steps and in a photon range between 25--36 eV, where the intensity of the bulk bands is substantially enhanced. Note that in this range we can follow the overall dispersion of the bulk bands because the bulk Brillouin zone is very small ($\mid$\kperp$\mid<$0.2 \invA). The reason is that the lattice constant of (Bi$_{1-x}$In$_x$)$_2$Se$_3$ is very large along the $z$-direction (according to our x-ray diffraction measurements, $c$ = 28.62\invA\ for Bi$_2$Se$_3$ which slightly increases linearly to $c$=28.63\invA\ for $x$=0.087). The resulting dispersions were then integrated over \kperp\ around \kpara=0. In Fig.~\ref{hvdep}(i), we clearly observe that the bulk band gap decreases continuously up to $x$=0.049. Above this value, the bulk band gap increases again, demonstrating that the transition point has been crossed and that the bulk band gap becomes positive. Subsequently, it reaches a value of $\sim$0.5 eV for $x$=0.087 which is larger than the bulk band gap in pure Bi$_2$Se$_3$. This behavior is fully consistent with the evolution of the bulk band gap as extracted from photoluminiscence experiments, as shown in Fig.~\ref{fig1expt}(g). 

\begin{figure}[tb!]
\centering
\includegraphics[angle=0,width=8.5cm]{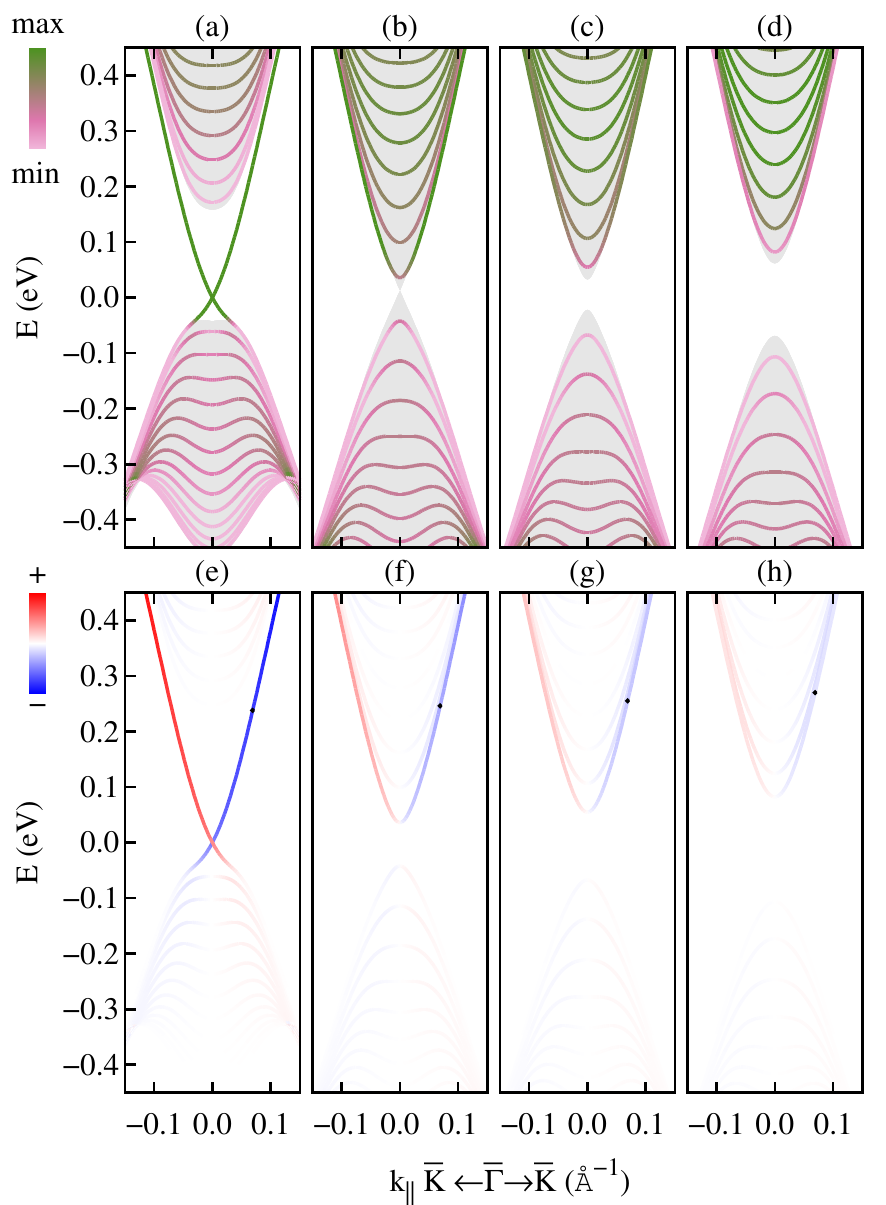}
\caption{\label{fig2bands} (Color online) Band structure of a 15-QL slab of Bi$_2$Se$_3$ obtained with a tight-binding model based on $GW$ with different SOC strengths: (a) and (e) 100\% SOC (non-trivial); (b) and (f) 77\% (non-trivial, very close to the transition); (c) and (g) 70\% (trivial); and (d) and (h) 60\% (trivial). The color scheme in panels (a)-(d) represents the localization of the states on the topmost QL. The bulk states are displayed as a gray background. The color in panels (e)-(h) shows the in-plane spin polarization, and the black dots represent the wave vector at which the spin polarization is discussed in Figs.~\ref{fig3spinpol} and~\ref{fig4depth} (\kpara\=0.07\AA$^{-1}$).}
\end{figure}

To identify the microscopic mechanisms underlying the appearance of gapped surface states, we performed $GW$-based tight-binding calculations of a slab of 15 QLs of \bise. For simplicity, in the calculations the quantum-phase transition is driven by varying the SOC strength in the absence of In. Figure~\ref{fig2bands} shows the electronic structure of bulk \bise~(gray background) and the 15-QL slab for selected SOC strengths in the non-trivial [Figs.~\ref{fig2bands}(a) and \ref{fig2bands}(b)] and trivial [Figs.~\ref{fig2bands}(c) and \ref{fig2bands}(d)] sides of the transition. The upper panels and their corresponding lower ones show the same electronic structure with a different color scheme: In Figs.~\ref{fig2bands}(a)-\ref{fig2bands}(d), the color scheme represents the localization of the states on the topmost QL. In Figs.~\ref{fig2bands}(e)-\ref{fig2bands}(h), the color intensity is proportional to the in-plane spin polarization. The calculations clearly show that with decreasing SOC, (i) a gap opens in the surface state on both sides of the transition, (ii) the gapped surface state survives the inversion of the bulk bands, (iii) the magnitude of the spin polarization progressively decreases and persists on the trivial phase. A summary of the results of the bulk (infinite) and surface (15 QLs) gaps are shown as red (dark) and green (light) solid lines in Fig.~\ref{fig1expt}(g), respectively.

Only on the basis of the $GW$ calculations, we derive that the SOC strength of pure In$_2$Se$_3$ corresponds to approximately 30\% of the SOC strength of pure Bi$_2$Se$_3$. We emphasize however that this correspondence cannot be directly used to estimate theoretically the critical $x$ value of the transition matching the band inversion. According to the $GW$ calculations, the transition occurs at 76\% SOC, which would correspond to a theoretical estimate of $x$=0.11. This concentration is overestimated with respect to the experimental observation by a factor of $\sim$2. The origin of this discrepancy is due to the fact that the In~5$s$ orbitals pull up in energy the valence-band maximum (VBM), increasing the value of the band gap in (Bi$_{1-x}$In$_x$)$_2$Se$_3$~\cite{brahlek2012,liu2013}. This effect promotes the topological phase transition making it to occur at a lower In content, in contrast to the case of e.g., (Bi$_{1-x}$Sb$_x$)$_2$Se$_3$, for which the transition critical concentration is almost uniquely determined by the SOC strength \cite{liu2013}. Equally important, calculations including substitutional In atoms of different concentrations explicitly~\cite{liu2013} also fail to reproduce the experimental value of the critical concentration, predicting approximately 17 mol.\%. 

The discrepancy between calculation and experiment affects the absolute correspondence between In content and SOC strength, but not our qualitative discussions. Therefore, in Fig.~\ref{fig1expt}(g) we used as an appropriate reference for this correspondence the transition critical point derived from calculations and experiments. In this way, we find that $x$=0.087 should correspond to about 50\% SOC, and the same trend follows through the whole quantum-phase transition. Such comparison indicates that in the experiment less than half of the In atoms do not occupy substitutional positions, but positions either in the vdW gaps or within the cation sublattice forming In$_2$Se$_3$ clusters, which is qualitatively in line with the results of our structural characterization in Figs.~\ref{fig1expt}(h)-~\ref{fig1expt}(l). Nevertheless, it should be emphasized that from this particular comparison the amount of non-substitional In atoms is quantitatively overestimated due to the effect of the In~5$s$ orbitals mentioned above. 

Because of the same reason, the quantitative agreement between the experimental and theoretical absolute values of the band gaps in Fig.~\ref{fig1expt}(g) should be also treated with caution. For example, the bulk band gap of pure $\alpha$-In$_2$Se$_3$ is larger than 1~eV~\cite{huiwen2013}, while \bise~with reduced SOC can maximally reach a bulk band gap of about 400~meV for 0\% SOC. 
Therefore, the absence of In atoms in our calculations makes the absolute values of the band gaps underestimated for a given SOC strength. Nevertheless, at a qualitative level the trends in the variation of the calculated bulk and surface band gaps are in good agreement with the tendencies revealed experimentally. 

\begin{figure}[tb!]
\centering
\includegraphics[width=0.47\textwidth]{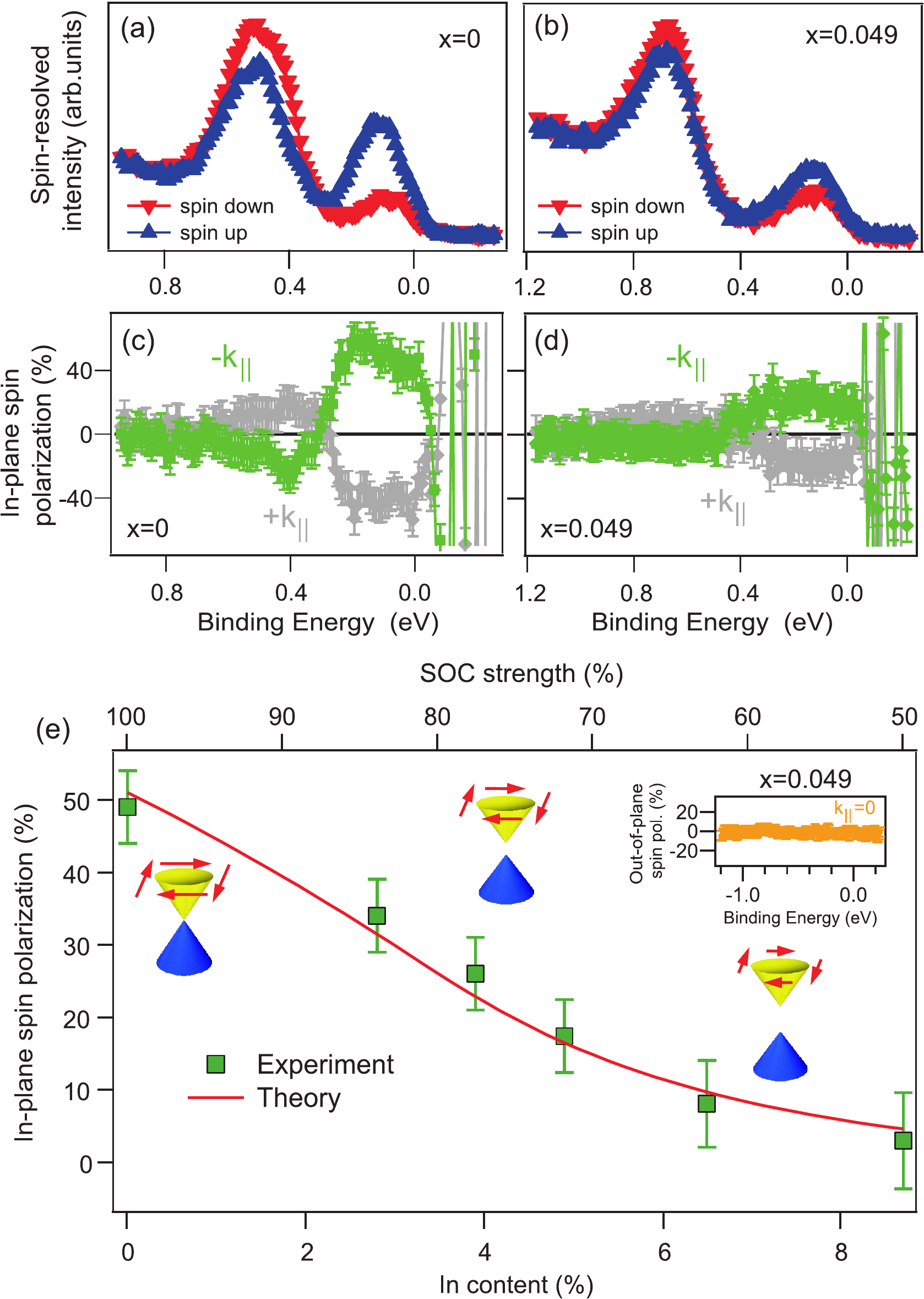}
\caption{\label{fig3spinpol} (Color online) (a)-(d) Selected spin-resolved ARPES measurements taken for In contents $x$ of (a),(c) 0 and (b),(d) 0.049 at 50 eV, revealing a decrease of the in-plane spin polarization. (a),(b) Spin-resolved EDCs obtained at \kpara=-0.07 \invA. The blue (red) curves show spin up (down) EDCs, and the orientation of the in-plane spin polarization is perpendicular to \kpara. (c),(d) Corresponding in-plane spin polarizations (green curves), which reverse at opposite momentum (gray curves). (e) Absolute magnitude of the in-plane spin polarization across the topological to trivial quantum-phase transition at \kpara=0.07 \invA. The green solid symbols are experimental data, and the red solid line the corresponding calculation. The TSS exhibits an unconventional transformation to a gapped surface state which is spin polarized and persists on the trivial phase. Top-right inset in (e): Out-plane spin polarization for $x$=0.049 at \kpara=0.}
\end{figure}

Specifically, whereas for 100\%~SOC no gap at the Dirac point of the TSS opens in the calculations, the reduction of the SOC~strength causes the appearance of gapped surface states. This result, which is in qualitative agreement with the data extracted from ARPES, will be discussed in more detail below. Furthermore, we observe a smooth closing of the bulk band gap across the phase transition upon SOC and $x$ variation in the theoretical and experimental results, respectively. This behavior appears in contrast to the scenario of a sudden closure of an otherwise constant bulk band gap as proposed in Ref.~\onlinecite{lou2015}. Conversely, our findings taken altogether clearly indicate that the gap closing is instead gradual and continuous. Although the origin of this discrepancy is not clear at the moment, we believe that it could be related to a misinterpretation in Ref.~\onlinecite{lou2015} of both the position of the VBM and the direct nature of the gap~\cite{nechaev2012}. In fact, the spin texture of the surface states exhibits very similar behavior, i.e., no sudden drop or discontinuous decrease of the spin polarization is found across the quantum-phase transition [Figs.~\ref{fig2bands}(e)-\ref{fig2bands}(h)]. Instead, a gradual, smooth decay of the spin polarization accompanies the transition, as summarized in Fig.~\ref{fig3spinpol}. Experimentally, this behavior is supported by our spin-resolved ARPES measurements taken at different \kpara\ wave vectors and as a function of In content. The data were acquired at 50 eV where the effect of light-induced manipulation of photoelectron spins does not contribute to the measured spin polarizations \cite{Sanchez-Barriga-PRX-2014}. 

In Figs.~\ref{fig3spinpol}(a)-\ref{fig3spinpol}(d) we show selected spin-resolved energy-distribution curves [Figs.~\ref{fig3spinpol}(a)-\ref{fig3spinpol}(b)] corresponding to the in-plane component of the surface-state spin polarization [Figs.~\ref{fig3spinpol}(c)-\ref{fig3spinpol}(d)], which clearly reverses at opposite \kpara\ wave vectors and decreases when In atoms are incorporated into the bulk. In Fig.~\ref{fig3spinpol}(e) we summarize the overall behavior of the magnitude of the in-plane spin polarization obtained from experiments and calculations as a function of In content and SOC strength, respectively. For Bi$_2$Se$_3$, the maximum observed in-plane spin polarization is about $(50 \pm 10)$\%. This value deviates from previous reports of nearly 100\% spin polarization from the TSS of \BiSe\ \cite{Pan11, Jozwiak11}. We attribute this deviation to the multiple orbital contributions from the TSS to the net spin polarization, as also derived from first-principles calculations on the same system \cite{Yazyev2010}.

The magnitude of the in-plane spin polarization progressively decreases both in experiment and calculations when moving from the topological to the trivial side of the quantum-phase transition. Moreover, there is a persistent in-plane spin polarization from the gapped surface states on the topologically-trivial phase in agreement with previous reports on BiTl(S$_{1-x}$Se$_{x}$)$_2$ \cite{xu2015}. In this respect, we emphasize that the calculated in-plane spin polarization becomes strictly zero only in the absolute absence of SOC. In contrast, the out-of-plane spin polarization was found to be negligible for all the states calculated in Fig.~\ref{fig2bands} and independent of SOC. Similarly, the measured out-of-plane spin polarization was found to contribute unobservably to the net spin polarization. The reason underlying this behavior is that the circular constant-energy contours of the surface states, as seen in Figs.~\ref{hvdep}(a) and ~\ref{hvdep}(b), do not contain contributions from hexagonal warping \cite{Fu-PRL-2009,Souma11}. 

It is also remarkable that for the gapped surface states the out-of-plane spin polarization remains zero at the \Gbar\ point of the SBZ [see inset in Fig.~\ref{fig3spinpol}(e)]. This behavior demonstrates the absence of a hedgehog spin texture which would be expected from broken time-reversal symmetry \cite{Sanchez16}. At first glance, such conclusion might seem evident because breaking time-reversal symmetry would require magnetic dopants. However, in the case of gapped graphene, the combination of spin-orbit coupling and sublattice asymmetry induces a hedgehog spin texture without the need of magnetic dopants \cite{Varykhalov2015}. Therefore, the lack of an out-of-plane spin polarization in our present case shows that, in the most general case, a hedgehog spin texture in TIs cannot be a straightforward consequence of the opening of a surface gap. 

\begin{figure}[tb!]
\centering
\includegraphics[angle=0,width=8.5cm]{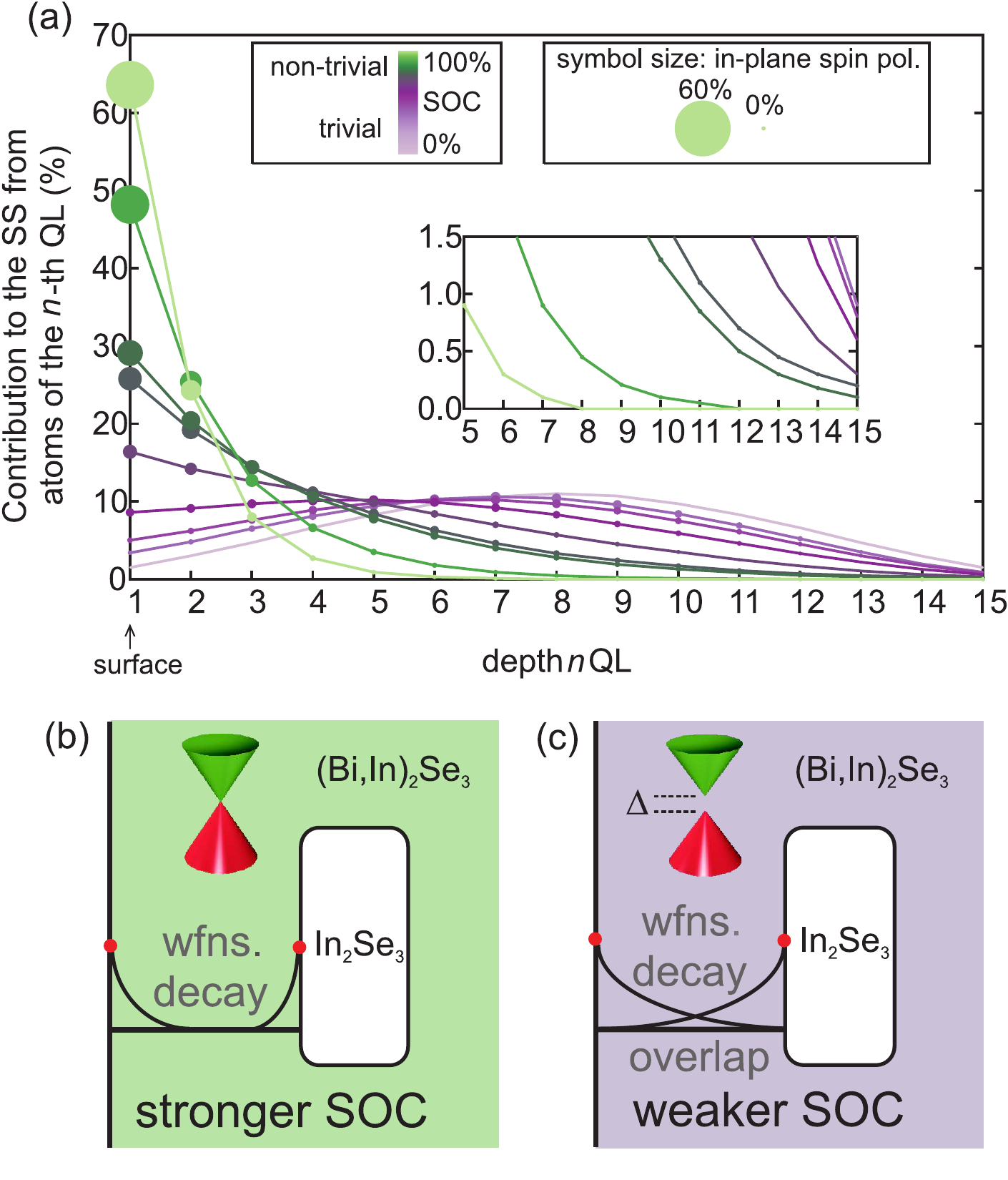}
\caption{\label{fig4depth} (Color online) (a) Wave functions (wfns.) decay in real space: Contribution from atoms of each QL (horizontal axis) to the wave function of the surface state at $\sim$0.07\AA$^{-1}$, corresponding to the black circles in Figs.~\ref{fig2bands}(e)-\ref{fig2bands}(h). The symbol size indicates the contribution from atoms of each QL to the in-plane spin polarization. Results are represented for different SOC strengths as indicated by the color scale: green, non-trivial; gray, transition; purple, trivial. Inset: Zoom-in on the lower right part of the figure. (b) and (c): Sketch of the proposed gapping mechanism on the topological side of the phase transition, depicting how trivial inhomogeneities are responsible for the opening of a surface gap (see text).}
\end{figure}

The band gap observed at the Dirac point of the topologically non-trivial system in Fig.~\ref{fig2bands}(b) is consistent with the low-frequency conductance measurements of Ref.~\onlinecite{wu2013}. To this end, the observed collapse in transport lifetimes across the quantum-phase transition in ultrathin (Bi$_{1-x}$In$_x$)$_2$Se$_3$ films was attributed to a gap opening at the Dirac point of the TSS \cite{wu2013}. This behavior can be in principle explained based on the ideal, general assumption that the penetration length of the TSS wave functions is inversely proportional to the bulk band gap~\cite{wu2013}. As seen in Fig.~\ref{fig4depth}(a), the gapped surface states on the topological side of the phase transition in our calculations of a 15-QL slab are caused by this effect. On the other hand, the opening of a surface gap on the trivial side of the phase transition is a natural consequence of the substantially reduced SOC~strength, both in our experiments and calculations. 

In particular, the opening of the surface gap on the topological side of the phase transition in our calculations is due to an unconventional cross-talk or hybridization mechanism between TSSs at opposite surfaces of the slab. This mechanism is unconventional because the cross-talk occurs at lower SOC strengths and at much higher thicknesses that previously reported for ultrathin TI films \cite{Neupane2014}. Based on the overall qualitative agreement between our calculations and experiments, we believe that a similar mechanism underlies the opening of the surface gap in our experiments, as we will discuss in detail below. In this respect, it should be mentioned that In 5$s$ and 5$p$ states cannot be the reason for the surface gap because they are strongly hybridized with host states \cite{liu2013}, and therefore do not show impurity-like character in contrast to the case of strongly $d$-like Mn states in Bi$_2$Se$_3$\cite{Sanchez16}. 

Let us firstly focus on our actual calculation of the decay of the wave functions of the surface states inside the bulk shown in Fig.~\ref{fig4depth}(a) for different SOC~strengths. This calculation was performed by projecting the wave functions of the surface states onto the maximally localized Wannier functions associated to the atoms in each QL. The SOC strength is indicated by the color of lines and symbols, and the spin polarization by the size of the symbols. There is a clear dominant contribution of the atoms of the first QL to the surface state even on the trivial side of the phase transition down to 70\% SOC, supporting the formation of a precursor state of the Dirac cone already on the trivial phase\cite{xu2015}. In contrast, below 70\% (i.e., away from the transition point) the curves become gradually much broader and the surface state progressively acquires more bulk-like character, reaching a regime in which it behaves as a surface resonance \cite{McRae1979}. In the topological side (green curves), the penetration depth of the surface state increases with decreasing SOC as a consequence of a larger bulk-surface interaction. A closer look at the values shows that only for 100\% and 90\% SOC, the surface state decays within the first 15~QLs. This behavior explains the gap opening at the Dirac point in our calculations of a 15-QL slab and provides clear insight on the mechanism underlying the behavior of the transport lifetimes across the quantum-phase transition~\cite{wu2013}.

Within the unconventional cross-talk picture revealed by our calculation in Fig.~\ref{fig4depth}(a), the spatial extent of the TSS is longer than half of the sample thickness such that the Dirac cone at the probed surface interacts with that at the opposite surface. This allows for the scattering of an electron in a state of one surface at $+$\kpara\ with another electron in a state of the other surface with the same spin at $-$\kpara, resulting in a gap opening. We propose that in our experiments, the gap opening obeys a similar mechanism, but the role of the second surface is now principally assumed by the In atoms in the cation sublattice, which form In$_2$Se$_3$ clusters as discussed above. Given the topologically trivial character of these clusters, there must necessarily exist a second Dirac cone at the interface between the non-trivial and trivial domains. Each of these Dirac cones would have a characteristic penetration depth away from the surface/interface depending on the coupling between surface and bulk states. The explanation that we propose for the appearance of the gap at the Dirac point of the TSS on the topological side of the phase transition is schematically shown in Figs.~\ref{fig4depth}(b) and \ref{fig4depth}(c): If the wave function of the Dirac cone at the surface that we probe decays slow enough to overlap with the wave function of the Dirac cone at the interface with a topologically-trivial region, the probed Dirac cone should present a gap, as we indeed observe in our experiments. In other words, if trivial inhomogeneities separated by an average distance $l$ appear in the interior of the sample, the observed Dirac cone behaves like that of an $l$-QL slab, irrespectively of the sample thickness. A key consequence is that the appearance of gapped surface states cannot be used as criterion to fully identify the trivial side of the quantum-phase transition in (Bi$_{1-x}$In$_x$)$_2$Se$_3$ bulk single crystals. 

Finally, we would like to address the role of small In concentrations as a decisive ingredient for the surface band gap opening on the topological side of the phase transition. If we would assume that about half of the In atoms occupy positions within the cation sublattice, $x$=0.028 would effectively correspond to about one In atom every ten inter-atomic distances (i.e., $\sim$ 4 nm). This value is considerably smaller than the penetration depth of the surface state even at 100\% SOC, as seen in Fig.~\ref{fig4depth}(a). It should be emphasized that this distance is underestimated due to the formation of In$_2$Se$_3$ clusters, which reduce the average density of In-related defects. The estimation is nevertheless not too far from an average cluster distance of $\sim$10-15~nm as inferred from our STM data in Fig.~\ref{fig1expt}(h). In addition, the overall effect could be electronically enhanced due to the long-range effective scattering potential in the surrounding of the clusters \cite{Black-Schaffer2012,Black-Schaffer2012a,Alpichshev2012}. In consequence, small In concentrations can have a remarkable impact on the energy-momentum dispersion of TSSs in (Bi$_{1-x}$In$_x$)$_2$Se$_3$ as observed in the present work.

\section{Conclusions}

We have examined the electronic structure of (Bi$_{1-x}$In$_x$)$_2$Se$_3$ single crystals across the topological to trivial quantum-phase transition and found that gapped surface states with non-zero spin polarization exist on both sides of the transition. We have found good qualitative agreement between our experimental results and relativistic many-body calculations upon variation of the In content and SOC~strength, respectively. This comparison clearly reveals the crucial impact of the spin-orbit interaction on the energy-momentum dispersion of the surface states as well as on the amplitude of bulk-to-surface coupling across the quantum-phase transition. To this end, we have provided an explanation to the behavior of the surface-state spin polarization and to the underlying mechanism giving rise to gapped surface states on the topological side of the phase transition. Our findings also reveal that a natural consequence of the reduction of the SOC~strength is the existence of gapped surface states with persistent spin polarization on the trivial side of the phase transition. We believe that the present results show a fundamentally distinct behavior of chemically-tuned quantum-phase transitions from topological to trivial insulators, solving the paradigm of an unconventional transformation of spin Dirac phases.

\begin{acknowledgments}

Financial support from SPP 1666 of the Deutsche Forschungsgemeinschaft and the Impuls-und Vernetzungsfonds der Helmholtz-Gemeinschaft (Grant No. HRJRG-408) is gratefully acknowledged. The theoretical work was supported by the Virtual Institute for Topological Insulators of the Helmholtz Association. We gratefully acknowledge computing time on the supercomputer JURECA at the J\"ulich Supercomputing Centre and the Russian-German bilateral program at BESSY-II. A.N.C. acknowledges support from the Russian Academy of Sciences and the Russian Foundation for Basic Research (Grant No. 17-02-01291). J.S.-B. and I. A. acknowledge helpful discussions with J. Min\'ar and C. Friedrich.

\end{acknowledgments}

\bibliographystyle{apsrev4-1}

\bibliography{references}

\end{document}